\documentclass{raa}
\usepackage{graphicx,times}
\usepackage{natbib}
\usepackage{amsmath}

\begin{document}

   \title{Detecting Radio Afterglows of Gamma-Ray Bursts with FAST $^*$
\footnotetext{\small $*$ Supported by the National Natural Science Foundation of China....}
}

 \volnopage{ {\bf 2014} Vol.\ {\bf X} No. {\bf XX}, 000--000}
   \setcounter{page}{1}

   \author{ Zhi-Bin Zhang\inst{1}, Si-Wei Kong\inst{2}, Yong-Feng Huang\inst{3}, Di Li\inst{4}, Long-Biao Li\inst{1}  }

   \institute{Department of Physics, College of Sciences, Guizhou
University, Guiyang 550025, China; {\it sci.zbzhang@gzu.edu.cn}\\
        \and
Key Laboratory for Particle Astrophysics, Institute of High Energy Physics, Chinese Academy of Sciences, 19B Yuquan Road, Beijing 100049, China \\
	\and
Department of Astronomy, Nanjing University, Nanjing 210093, China; {\it hyf@nju.edu.cn}\\
\and
National Astronomical Observatories of China, Chinese Academy of Sciences, 20A Datun Road, Beijing 100020, China
\\
\vs \no
   {\small Received 2014 Feb 24; accepted 2014 May 27}
}

\abstract{Using the generic hydrodynamic model of Gamma-Ray Burst (GRB) afterglows, we calculate the radio afterglow light curves of low luminosity, high luminosity, failed and standard GRBs in different observational bands of FAST's energy window. The GRBs are assumed to be located at different distances from us. Our results rank the detectability of GRBs in descending order as high luminosity, standard, failed and low luminosity GRBs. We predict that almost all types of radio afterglows except that of low luminosity GRBs could be observed by the large radio telescope as long as the domains of time and frequency are appropriate. It is important to note that FAST can detect relatively weak radio afterglows at a higher frequency of 2.5 GHz for very high redshift up to z=15 or even more. Radio afterglows of low luminosity GRBs can be detected only after the completion of the 2nd phase of FAST. FAST is expected to significantly expand the sample of GRB radio afterglows in the near future.
\keywords{gamma rays: bursts --methods: numerical--telescopes }
}

   \authorrunning{Zhang, Kong, Huang et al.}            
   \titlerunning{Gamma-Ray Burst Radio Afterglows}  
   \maketitle

%
\section{Introduction}           
\label{sect:intro}
With prompt developments of space and ground telescopes, studies of Gamma-Ray Bursts (GRBs) have come into an era of full wavelengths (e.g. Gehrels \& Razzaque 2013). Particularly, with recent space missions such as Swift satellite, the rapid response together with accurate localization enables more detailed follow-up observations by the ground facilities at longer times and lower frequencies. It has led to deeper understanding of the physical origins of GRBs. In addition, statistical analysis becomes possible and it is very helpful to compare the properties of GRBs and their afterglows (Sakamoto et al. 2011). Prompt $\gamma$-rays and their follow-up X-ray and Optical afterglows are found to be correlated with each other, more or less, for both short and long-duration bursts (Gehrels et al. 2008; Nysewander et al. 2009; Kann et al. 2011). It is interesting that Chandra \& Frail (2012) found the detectability of radio afterglows to be correlated with neither the $\gamma$-ray fluences nor the X-ray fluxes, but only with optical brightness in statistics. However, observational constraints of radio afterglows are relatively insufficient, although some authors have recently compiled larger datasets (Postigo, 2012; Chandra \& Frail 2012; Ghirlanda et al. 2013; Staley et al. 2013) since the first radio afterglow of GRB 970508 was discovered (Frail et al. 1997). We need to synthesize the GRBs and their afterglows at multiple energy bands separately and then combine them to explore their comprehensive physics.

 In the framework of the fireball internal-external shock model, GRBs are produced when the kinetic energy of an ultra-relativistic flow is dissipated by internal collisions, while the afterglows are emitted when the flow is slowed down by external shocks with the surrounding matter of the burst. The fireball model has given numerous successful predictions on GRB afterglows, such as the afterglow itself, jet break in the light curve of afterglows, the optical flash and the afterglow shallow-decay phenomena, etc. (see Lu et al. 2004, Piran 2004, Zhang 2007 and \textbf{Gao et al. 2013} for a review). Huang et al. (1999, 2000b) proposed a set of simplified dynamical equations that is consistent with the self-similar solution of Blandford \& McKee (1976) in the ultra-relativistic phase, and also consort with the Sedov solution (Sedov 1969) in the non-relativistic phase. Therefore, these equations can conveniently describe GRB afterglows at all post-burst times. For instance, the beaming effects (Rhoads 1997, 1999; Huang, Dai \& Lu 2000c), the rebrightening at multiple-wavelengths(Huang et al. 2004; Xu \& Huang 2010; Kong, Wong, Huang \& Cheng 2010; Yu \& Huang  2013) and the multi-band afterglow modeling (Huang, Dai \& Lu 2000c; Huang, Dai \& Lu 2002; Huang \& Cheng 2003; Wang, Huang \& Kong 2009; Kong, Huang, Cheng \& Lu 2009) can be easily dealt with by these equations.

 Square Kilometer Array (SKA) will be the largest and most sensitive radio telescope group in the world. The SKA project is designed to be constructed via two phases and will receive radio signals at a frequency range from 70 MHz to 10 GHz. As the largest worldwide single-dish radio telescope, the Five-hundred-meter Aperture Spherical radio Telescope (FAST, Nan et al. 2011; Li, Nan \& Pan 2013) is a Chinese megascience project that is being built in Guizhou province of southwestern China with an expected first light in Sep. of 2016. FAST continuously covers radio frequencies between 70 MHz and 3 GHz. A possible extension to 8GHz is being considered in the 2nd phase of FAST. FAST will be equipped with a variety of instruments and has been designed for different scientific purposes including the radio afterglows of GRBs. According to the current data sets presented by Chandra \& Frail (2012), the detection rate of radio afterglows is 30 \%, in which more than half of the radio flux measurements are made at 8.5 GHz. However, nearly 10 percent of the detected radio afterglows are from bright long GRBs (Ghirland et al. 2013; Salvaterra et al. 2012). The reason is that the low-frequency observations could be more affected by the bias of receivers.

In this work, we apply the fireball model to calculate a variety of numerical afterglow curves with changing redshifts for different cases representing failed, low luminosity, high luminosity and standard GRBs, respectively. These theoretical light curves based on diverse physical considerations are directly compared with FAST's sensitivity in order to diagnose the detectability by FAST. Our radio light curves within the FAST's window are derived for both low (70 MHz-0.5 GHz) and medium/high frequencies (0.5-3 GHz). The structure of our paper is as follows. Firstly, we provide an overview on observations of GRB radio afterglows in Section 2. Theoretical dynamical model of afterglow and sensitivity of FAST are introduced in Section 3. Numerical results are shown in Section 4 and we end with discussions and brief conclusions in Section 5.

\section{Overview of radio afterglow observations}
Recently, Chandra \& Frail (2012) presented a large sample of GRB radio observations for a 14-year period since 1997. Despite 304 radio afterglows consisting of 2995 flux measurements, only 95 out of 304 were reported to have radio afterglows detected by VLA, corresponding to a detection rate of $\sim$30\%, of which 1539 measurements are made in 8.5 GHz. They pointed out that the current detection rate of radio afterglows, much lower than in the X-ray (~90\%) or optical (~75\%) bands in the Swift era, may be seriously limited by instrumental sensitivity. Hancock, Gaensler \& Murphy (2013) argued that the lower rate would be caused by two intrinsically different types of bursts, namely radio bright and radio faint sources. However, radio emissions from host galaxies of GRBs also make the radio afterglows more difficult to detect (e.g. Berger, Kulkarni \& Frail 2001; Berger 2014; Li et al. 2014). The Chandra \& Frail sample contains 33 short-hard bursts, 19 X-ray flashes, 26 GRBs/SNe candidates including low luminosity bursts and 4 high-reshift bursts, of which only a few radio afterglows are available owing to their lower detection rate. In general, short bursts with smaller istropic energy, similar to low luminosity ones, are thought to occur in a relatively neaby universe like supernova. On the other hand, afterglows of short bursts are usually much dimmer than those of long ones (see e.g. Rowlinson 2013 for a review). This motivates us to focus on contrasting the radio afterglows in Fig. 1 between short GRBs (050724, 051221A and 130603B), low luminosity GRBs (060218) and high-redshift GRBs (050904, 080913, 090423 and 090429B) with successful detections. Note that only GRB 130603B was not included in Chandra \& Frail (2012) and its data are taken from Fong, Berger \& Metzger et al (2014). As the largest next-generation single dish radio telescope, FAST is expected to bring us many important findings (Nan, Li \& Jin et al. 2011; Li, Nan \& Pan 2013). The FAST's upper flux limits (see below) of detecting these kinds of bursts are also given in Fig. 1, showing that radio afterglows of the above-mentioned three kinds of special bursts would be easily detected by FAST. Interestingly, it is noted that high-redshift bursts, similar to short bursts except GRB 050724, have typical flux density less than 150 $\mu$Jy in radio band. Considering the above situations, we shall simulate theoretical radio afterglows of diffent types of bursts and probe the detectablity of FAST to them subsequently.

\begin{figure}
   \centering
 \includegraphics[width=14.5cm, height=12.5cm, angle=0]{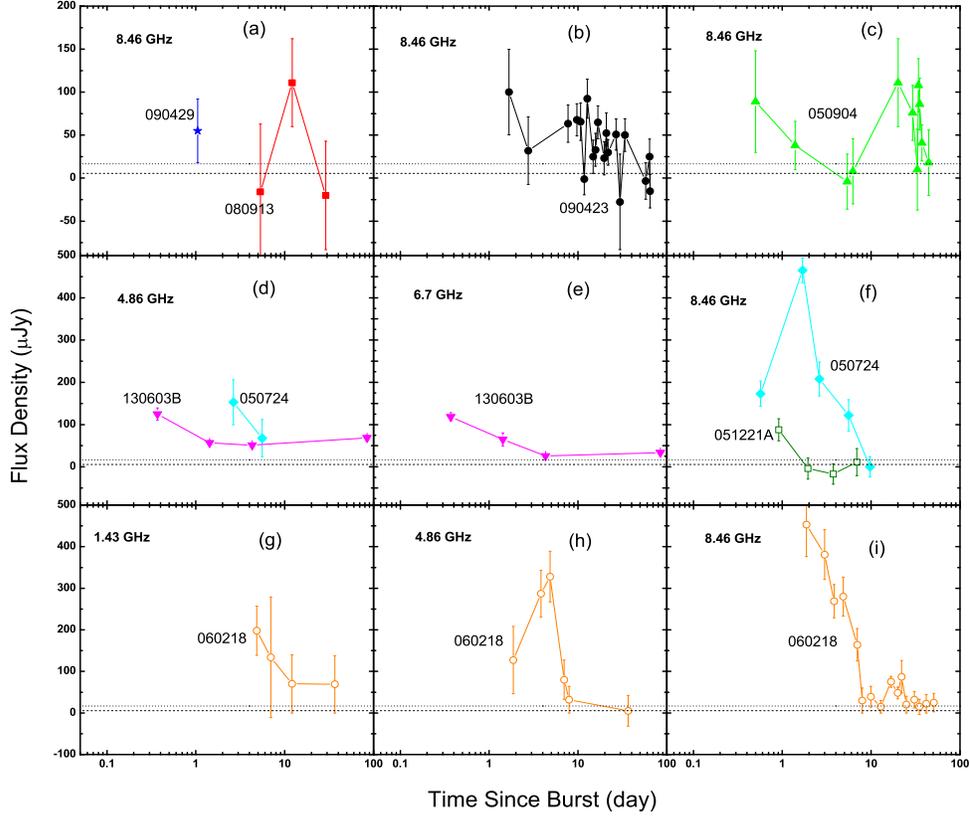}
   \caption{Broad-band radio afterglows of high-redshift bursts (star: GRB 090429B; filled-squares: GRB 080913; filled-circles: GRB 090423; filled-triangles: GRB 050904) in panels (a)-(c), short bursts (filled-triangles: GRB 130603B; filled-diamonds: GRB 050724; empty squares: GRB 051221A) in panels (d)-(f) and SNe-associated bursts (empty circles: GRB 060218) in panels (g)-(i). The dotted and dashed lines respectively represent $3\sigma$ and $1\sigma$ sensitivity limit of FAST in an integral time of 10 minutes. Note that the data of GRB 130603B are taken from Fong et al. (2014) and\textbf{ all the other radio data are collected from http://heasarc.gsfc.nasa.gov/W3Browse/all/rssgrbag.html (Chandra \& Frail 2012) directly.}}
   \label{Fig1}
   \end{figure}

\section{Dynamical Model and FAST Sensitivity}
\label{sect:Obs}

In terms of the generic dynamical model (Huang, Dai \& Lu 1999, 2000a, 2000c), the overall evolution of the ejected outflows in either ultra-relativistic or non-relativistic (Newtonian) phase can be generally described as

\begin{equation}
\frac{dR}{dt}=\beta c\gamma(\gamma+\sqrt{\gamma^2-1}),
\end{equation}
\begin{equation}
\frac{dm}{dR}=2\pi(1-cos\theta)R^{2}nm_p,
\end{equation}
\begin{equation}
\frac{d\theta}{dt}=\frac{c_s}{R}(\gamma+\sqrt{\gamma^2-1}),
\end{equation}
\begin{equation}
\frac{d\gamma}{dm}=-\frac{\gamma^2-1}{M_{ej}+\varepsilon m+2(1-\varepsilon)\gamma m},
\end{equation}
where $R$ is the radial distance measured in the source frame from the initiation point; $m$ is the rest mass of the swept-up circumburst medium; $\theta$ is the half-opening angle of the ejecta; $\gamma$ is the bulk Lorentz factor of the moving material; $t$ is the arrival time of photons measured in the observer frame; $\beta=\sqrt{1-\gamma^{-2}}$, $c$ is the speed of light. $n$ is the number density of Interstellar Medium; $\varepsilon$ is the general radiative efficiency and would evolve from 1 (high radiative case) to 0 (adiabatic case) within several hours after a burst; $M_{ej}$ is the initial rest mass of the ejecta. $c_s$ is the comoving sound speed determined by $c_s^{2}=\hat{\gamma}(\hat{\gamma}-1)(\gamma-1)c^2/[1+\hat{\gamma}(\gamma-1)]$ with the adiabatic index $\hat{\gamma}=4/3$ in the ultra-relativistic limit and $\hat{\gamma}=1$ in the Newtonian limit (Huang et al. 2000b). We further define $\xi_e$ and $\xi_B^2$ as the energy equipartition factors for electrons and the comoving magnetic field, respectively.

Denoting $\Theta$ as the viewing angle between the velocity of ejecta and the line of sight and $\mu=cos\Theta$, in the burst comoving frame, we can obtain synchrotron radiation power of electrons at frequency $\nu '$ as (Rybicki \& Lightman 1979)
\begin{equation}
P'(\nu')=\frac{\sqrt{3}e^3B'}{m_ec^2}\int_{min(\gamma_{e,min}, \gamma_c)}^{\gamma_{e,max}}(\frac{dN_e^{'}}{d\gamma_e})F(\frac{\nu'}{\nu_c^{'}})d\gamma_e,
\end{equation}
in which $dN_e^{'}/d\gamma_e\propto(\gamma_e-1)^{-p}$, with the typical value of the electron distribution index $p$ between 2 and 3 (Huang \& Cheng 2003); $\nu_c^{'}=3\gamma_e^2eB'(4\pi m_ec)^{-1}$ is the characteristic frequency of electrons with charge $e$; $\gamma_c=6\pi m_ec/(\sigma_T\gamma B'^2t)$ is the typical Lorentz factor of electrons which cool rapidly due to synchrotron radiation, with $\sigma_T$ being the Thompson cross-section; $\gamma_{e,min}=\xi_e(\gamma-1)m_p(p-2)/[m_e(p-1)]+1$ and $\gamma_{e,max}\simeq10^8(B'/1G)^{-1/2}$ are respectively the minimum and the maximum of Lorentz factors of electrons, and $F(x)=x\int_x^{+\infty}K_{5/3}(x')dx'$ of which $K_{5/3}(x')$ is the Bessel function. Owing to cosmological expansion, the observed frequency $\nu$ should be (Wang, Huang \& Kong 2009)
\begin{equation}
\nu=[\gamma(1-\beta\mu)]^{-1}\nu'/(1+z)=\delta\nu'/(1+z),
\end{equation}
here $\delta=[\gamma(1-\beta\mu)]^{-1}$ is the Doppler factor. For low-frequency radiation, the effect of synchrotron self-absorption on observation should be considered and hence the observed flux density radiated from a cosmological point source would be (Huang et al. 2000b; Wang, Huang \& Kong 2009)
\begin{equation}
F_{\nu}(t)=\frac{(1+z)\delta^3}{4\pi D_L^2}f(\tau)P'[ (1+z)\nu/\delta],
\end{equation}
where $D_L$ is the luminosity distance and $f(\tau)=(1-e^{-\tau_{\nu'}})/\tau_{\nu'}$ is a reduction factor of synchrotron self-absorption with an optical depth $\tau_{\nu'}$. In order to calculate the total observed flux densities, we should integrate Eq. (7) over the equal arrival time surface determined by
\begin{equation}
t=(1+z)\int\frac{1-\beta\mu}{\beta c}=const,
\end{equation}
within the jet boundaries.

Basically, the above model can give a good description for the external shocks and GRB afterglows. For example, the dynamical model has been applied to many GRBs and can well explain the observations (Huang, Cheng \& Gao 2006; Kong, Huang \& Cheng 2009; Xu, Huang \& Lu 2009; Kong, Wong, Huang \& Cheng 2010; Xu, Nagataki \& Huang 2011; Yu \& Huang 2013; Geng, Wu, Huang et al. 2013). Here in Fig. 2, we display some exemplar afterglow light curves at X-ray band (0.3-10 keV) and optical R-band calculated by using this model. It can be seen that this model basically matches the general behaviors of multi-band afterglows (e.g., Zhang et al. 2006; Nousek et al. 2006; Zhang, Liang \& Zhang 2007; Liang et al. 2008; Troja et al. 2007). Besides, afterglows from the extreme GRBs such as high luminosity, low luminosity and failed bursts are also predicted to shed new light on the future observations in the era of FAST.
\begin{figure}
   \centering
 \includegraphics[width=12cm, height=10cm, angle=0]{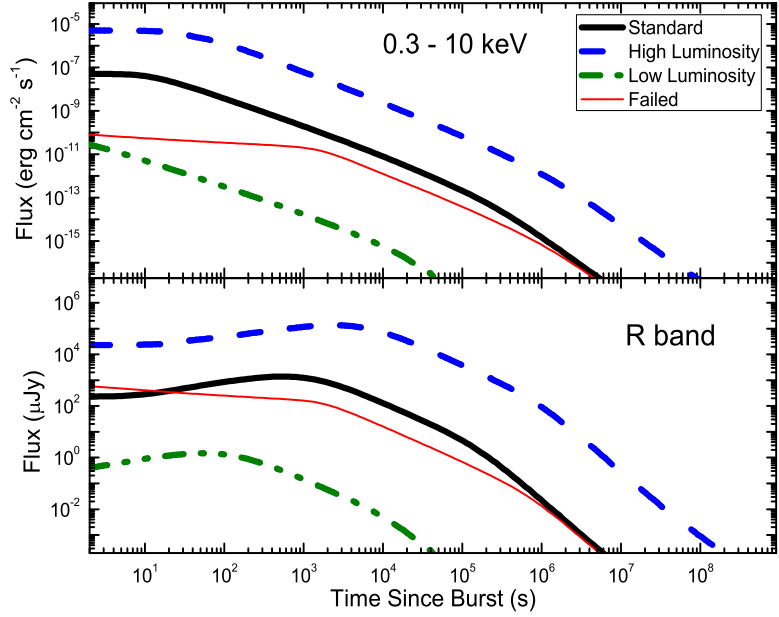}
   \caption{Plots of afterglow light curves for the standard (thick solid line), high luminosity (dashed line), low luminosity (dash-dot-dotted line) and failed (thin solid line) GRBs at X-ray energy band (0.3-10 keV) in upper panel and optical R band in lower panel, respectively, for a typical redshift z=0.5 and density n=1 $cm^{-3}$. See the text for more details.}
   \label{Fig2}
   \end{figure}

In order to explore the detectability of FAST, we here adopt the following sensitivity to estimate the RMS noise in a position-switching observing mode as

\begin{equation}
F_{lim}=\frac{(S/N)2 K_b T_{sys}}{A_e \sqrt{\Delta\tau \Delta \nu}}\simeq (12\mu Jy)(\frac{0.77\times10^3 m^2/K}{A_{e}/T_{sys}})(\frac{S/N}{3})(\frac{1 hour}{\Delta \tau})^{1/2}(\frac{100 MHz}{\Delta \nu})^{1/2},
\end{equation}
where $A_e$ is the effective area defined by $A_e=\eta_A A_g$ with an aperture efficiency $\eta_A=0.65$ (Yue et al. 2013), $\Delta\tau$ and $\Delta\nu$ are respectively the bandwidth and the integral time, the illuminated geometric area is $A_g=\pi\times(300/2)^2$ m$^2$, and the system temperature is $T_{sys}$ for the FAST. Table~\ref{tab1} gives the values of sensitivity for different frequencies of FAST. $S/N$ stands for the signal-to-noise ratio and is generally taken as no less than 3. More detailed properties of FAST can be found in Nan et al. (2011). Note that we have calculated FAST's sensitivities by assuming that its temperature template is similar to that of VLA's receivers. We acknowledge here the difficulty of the bandpass calibration for a single dish, which is more difficult than that for interferometers. The detection of faint, broadband continuum signal, however, has been achieved by single dishes before. FAST is also looking into new calibration technologies, such as injecting an artificial flat spectrum signal into the optical path.

\begin{table}
\bc
\begin{minipage}[]{300mm}
\caption[]{Parameters of the Nine Sets of FAST's Receivers\label{tab1}}\end{minipage}
\setlength{\tabcolsep}{6pt}
\small
 \begin{tabular}{|c|c|c|c|c|c|c|c|c|c|c|c|}
  \hline\noalign{\smallskip}
 No.&Bands$^{\dag}$&$\nu_c^{\dag}$& $\Delta\nu^{\dag}$& $T_{sys}^{\dag}$  & $F_{lim,1}$& $F_{lim,2}$& $F_{lim,3}$& $T_{sys}'$& $F_{lim,1}'$& $F_{lim,2}'$& $F_{lim,3}'$\\
&$GHz$&$GHz$&$MHz$&$K$&$\mu Jy$&$\mu Jy$&$\mu Jy$&$K$&$\mu Jy$&$\mu Jy$&$\mu Jy$\\
  \hline\noalign{\smallskip}
1&0.07-0.14&0.1&70&1000&293.6&168.7&119.4&140&41.0&23.5&16.9\\
\hline
2&0.14-0.28&0.2&140&400&82.5&47.6&33.8&140&28.8&16.6&11.9\\
\hline
3&0.28-0.56&0.4&280&150&21.9&12.7&8.9&130&19.1&11.1&7.8\\
\hline
4&0.56-1.02&0.8&560&60&6.1&3.6&2.5&97&9.9&5.8&4.2\\
\hline
5&0.32-0.334&0.328&14&200&130.7&75.3&53.5&125&81.7&47.1&33.2\\
\hline
6&0.55-0.64&0.6&90&60&15.5&8.9&6.4&112&28.9&16.6&11.9\\
\hline
7&1.15-1.72&1.45&550&25&2.5&1.4&1.1&61&6.3&3.6&2.5\\
\hline
8&1.23-1.53&1.38&300&25&3.6&1.9&1.4&60&8.6&5.0&3.6\\
\hline
9&2.00-3.00&2.5&1000&25&1.9&1.1&0.8&30&2.2&1.4&0.8\\
  \noalign{\smallskip}\hline
\end{tabular}
\ec
\tablecomments{0.86\textwidth}{$^{\dag}$ These data are taken from Nan et al. (2011). $F_{lim,i}$ ($i$=1, 2, 3) denotes the limited flux densities for different integral time $\Delta\tau$ equaling 10, 30 and 60 minutes. The limited flux densities $F_{lim,i}'$ ($i$=1, 2, 3) are estimated in terms of the system temperatures of the VLA's template (https://science.nrao.edu/facilities/vla) within the above corresponding time levels.}
\end{table}

\section{Results}
\label{sect:Results}
To determine the sensitivity of FAST to GRB radio afterglows, we have chosen the standard, high-luminosity, low-luminosity and failed GRBs for a comparative study. For convenience, let us define the initial values or parameters of these bursts as follows. (1) Standard GRBs: initial isotropic energy $E_0=10^{52}$ ergs and $\gamma_0=300$; (2) High luminosity GRBs:  $E_0=10^{54}$ ergs and $\gamma_0=300$; (3) Low luminosity GRBs: $E_0=10^{49}$ ergs and $\gamma_0=300$; (4) Failed GRBs: $E_0=10^{52}$ ergs and $\gamma_0=30$. Apart from these differences, we assume all the bursts have the same values for other parameters, namely $n=1$ cm$^{-3}$, $p=2.5$, $\xi_e=0.1$, $\xi_B^2=0.001$, $\theta=0.1$ and $\Theta=0$ throughout this paper. Also, an assumption of the radiative efficiency $\varepsilon=0$ has been made because the relativistic fireball becomes fully adiabatic in about several hours after a burst. Light-curves of radio afterglows at different redshifts, i.e. $z$=0.5, 1, 5, 10, 15, have been derived in order to study FAST's capability of probing bursts in the early universe.

\subsection{Light-Curves of Radio Afterglows in the FAST Window}
Figs. 3-7 show that almost all radio afterglow light curves, except the failed GRB afterglows have the same general characteristics of slow rise and fast decay (SRFD) although they may have distinct physical origins. We also find a common and interesting phenomenon that with the increase of observing frequency, the light curve peaks earlier, and the peak flux is also higher. At the same time, radio afterglows of the standard and the failed GRBs have almost the same peak time and the same peak flux density due to their similar kinetic energies. The two kinds of bursts decay congruously after their peak times. At higher frequencies and larger redshifts, the peak flux densities of failed GRBs are slightly weaker than those of the standard ones. Both the peak flux density and the peak time of radio afterglows sensitively depend on the initial energy injection. For the failed GRBs, the rising part is largely affected by the small initial Lorentz factors of the ejecta from the central engine. In any cases, we notice that radio afterglows of the low luminosity GRBs have the lowest brightness except that they are stronger than those of the failed GRBs at early stage of less than 1hr. The intersection point would be postponed when the observing frequency is relatively lower for a farther burst. Note that it seems unlikely for FAST to detect any kinds of radio afterglow emission at the extremely lower frequency of $\leq$ 0.1 GHz. For bursts at the same distance or redshift, their radio flux density in higher frequency is usually stronger than that in lower frequency when they are observed at the same time and could  be detected at a very early stage. Furthermore, it is found that the radio flux densities are relatively insensitive to the redshift as seen in Figs 5-7, which is consistent with previous investigations (e.g. Ciardi \& Loeb 2000; Gou et al. 2004; Frail et al. 2006; Chandra \& Frail 2012).


\subsubsection{Standard GRBs}
The thick solid lines in Figs. 3-7 denote the case of a standard fireball in different energy bands and redshifts. For the redshift $z=0.5$, the peaked radio emissions at $\nu>$ 0.4 GHz can be easily detected by FAST. With the increase of observing frequency, the bursts gradually brighten. FAST can even detect very early radio afterglows in the prompt phase for a time of less than 10 second at 2.5 GHz. It is interesting that in high frequency bands, the radio afterglow can typically be observable for $\sim115$ days. The peak flux densities at 1.4 GHz and 2.5 GHz can reach 70 $\mu$Jy and 200 $\mu$Jy, respectively. For the reshift $z=1$, FAST can detect the radio emission in the frequency band of $\nu>$0.6 GHz, especially $\nu=$2.5 GHz from 20 seconds to 62 days since a burst for 1 hour integration time. The earliest detection time may start as early as 10 seconds. For the redshift $z=5$, radio afterglows under 0.8 GHz are undetectable for FAST. The peak flux densities at 0.8 GHz and 2.5 GHz can respectively reach 3 $\mu$Jy and 6 $\mu$Jy, which will be observed up to 52 days from the earliest starting time of 30 minutes. For much higher redshifts of $z=10$ and 15, the radio flux densities have a peak value of $\sim2\mu$Jy and only radio afterglows above 1.4 GHz can be marginally detected. Note that the GRB radio afterglows in the standard case usually peak at 10-100 days, which is well consistent with the observations described in Chandra \& Frail (2012).

\subsubsection{High Luminosity GRBs}

The high luminosity GRBs marked with thick dashed lines in Figs. 3-7 are driven by the largest energy ejection and their radio afterglows naturally hold the strongest brightness when they are located at the same distance. For nearby bursts with a redshift of $z=0.5$, FAST has the capability of detecting all radio afterglows at a frequency of no less than 100 MHz. The typical peak flux densities are 100 $\mu$Jy  at 0.2 GHz and $7\times10^3\mu$Jy at 2.5 GHz respectively. The former can be detected from 3 days to 3.5 years and the later can be observed from several seconds to 9 years after the burst. For a redshift $z=1$, FAST may detect the weak peak flux in the post-burst time of 2 hours to 1200 days in channel 3 at 0.4 GHz and 30 seconds to 6 years in a frequency of $\nu=1.4$ GHz. The radio afterglow flux densities for redshift higher than 5 are undetectable under $\nu\simeq$300 MHz but can be safely detected from 1 day to 2.5 years at 0.4 GHz and from several tens seconds to 4 years in high frequency bands. The peak radio flux densities in 0.4 -2.5 GHz for $z=15$ can reach 50-200 $\mu$Jy which is much higher than the threshold of FAST.

\subsubsection{Low Luminosity GRBs}

In contrast, the low luminosity GRBs denoted by dash-dot-dot lines in Figs. 3-7 consist of some special bursts with lower energy input. This leads to smaller kinetic energy of outflows and then much weaker radio afterglows peaking at earlier time around 1 day after the GRB trigger. The weakest radio brightness is only $\sim10^{-4}\mu$Jy at 100 MHz for redshift $z=15$. The strongest flux density can approach $0.3\mu$Jy at 2.5 GHz for $z=0.5$ and is just close to the detection limit of FAST for 1 hour integration time. It is obvious that FAST can hardly detect radio afterglows from these low luminosity GRBs. In the future, if FAST's passband can be expanded to 8 GHz, then low luminosity GRBs may also be detected. We can also consider to increase the integration time to tell apart the radio afterglow from a low level background.

\subsubsection{Failed GRBs}
Such kinds of bursts are thought to be produced by an isotropic fireball with kinetic energy like that of the standard bursts but with much lower Lorentz factors of several tens. As shown in Fig. 3-7, radio afterglows of failed GRBs and the standard ones nearly peak simultaneously and they decay with time in the same way after the peak time. Compared with standard bursts, radio afterglows of failed bursts can be detected mainly at later times. For $z=0.5$, FAST can hardly detect them at a frequency lower than 0.6 GHz in 3-$\sigma$ levels with 1 hour integration time. But, we can detect the radio afterglows up to 85 days at higher frequencies above $0.8$ GHz. The afterglow is observable from a time of 1000 minutes, 70 minutes, 20 minutes, 15 minutes and 8 minutes for 0.6 GHz, 0.8 GHz, 1.38 GHz, 1.45 GHz and 2.5 GHz after the burst, respectively. For the redshift $z$=1, only radio emissions at a frequency larger than 0.6 GHz are detectable from 1 day to 60 days. Radio afterglows at a redshift $z=$5 for a frequency larger than 1.4 GHz can be observed. The radio flux densities at 0.8 GHz, 1.38 GHz, 1.45 GHz and 2.5 GHz are in a lower level of 1-2 $\mu$Jy. For $z>10$, radio afterglows of the failed GRBs become very difficult to be detected by FAST with its current sensitivity of 1 hour integration time.

\begin{figure}
   \centering
 \includegraphics[width=16cm, height=14cm, angle=0]{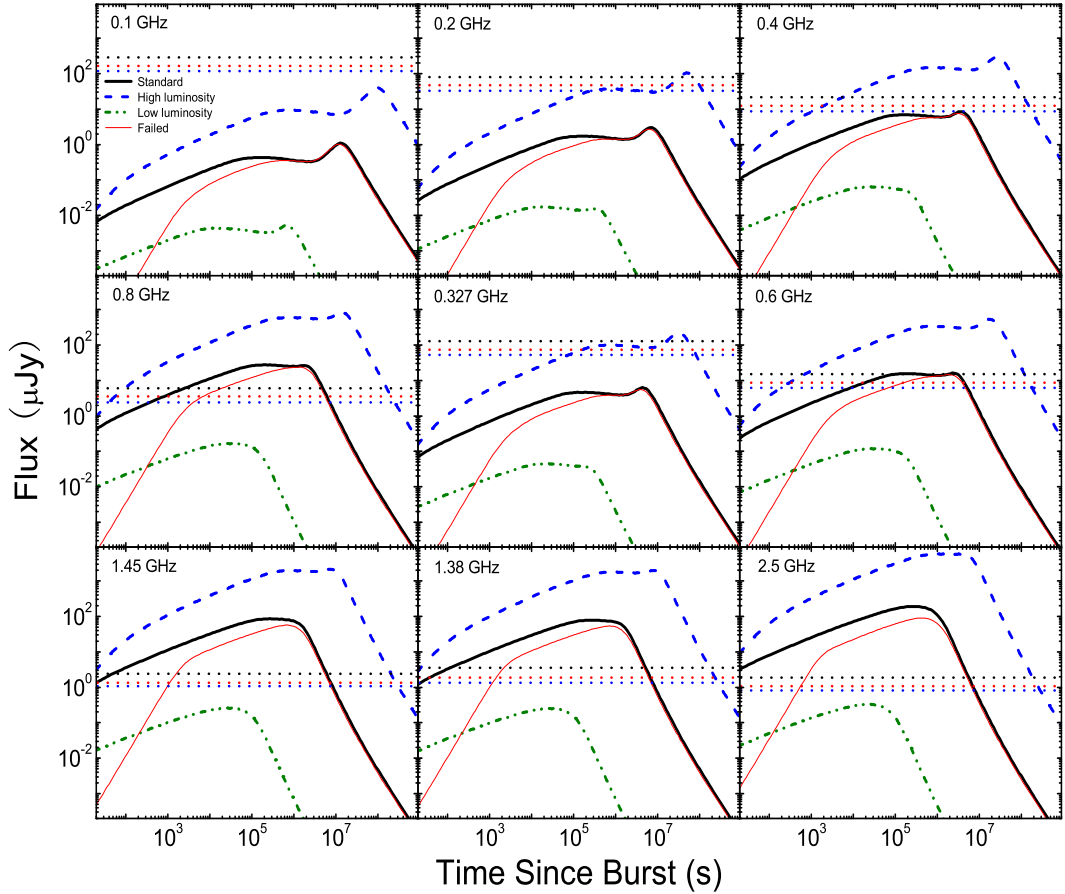}
   \caption{Radio flux density of GRBs at a redshift $z=0.5$ versus observation time $t$ in the observer frame at various frequencies within the FAST's window. The radio light curves of standard, high-luminosity, low-luminosity and failed GRBs are marked with thick solid, dashed, dash-dot-dot and thin solid lines, respectively, and have been symbolized in panel 1. Three horizontal dotted lines from upper to bottom represent 1$\sigma$ ($S/N$=1) limiting flux density of the FAST for 10, 30 and 60 minutes integration time correspondingly. See the text for details.}
   \label{Fig3}
   \end{figure}

   \begin{figure}
   \centering
 \includegraphics[width=14.5cm, height=12cm, angle=0]{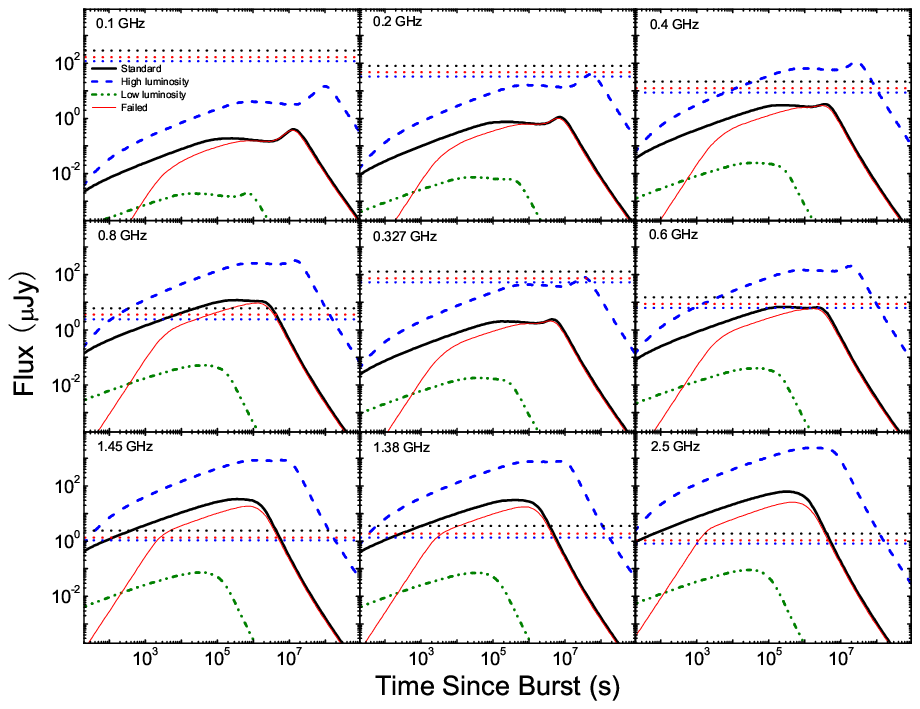}
   \caption{Radio flux density of GRBs at a redshift $z=1$ versus observation time $t$ in the observer frame at various frequencies within the FAST's window. Symbols are the same as in Fig. 1.}
   \label{Fig4}
   \end{figure}

   \begin{figure}
   \centering
\includegraphics[width=14.5cm, height=12cm,angle=0]{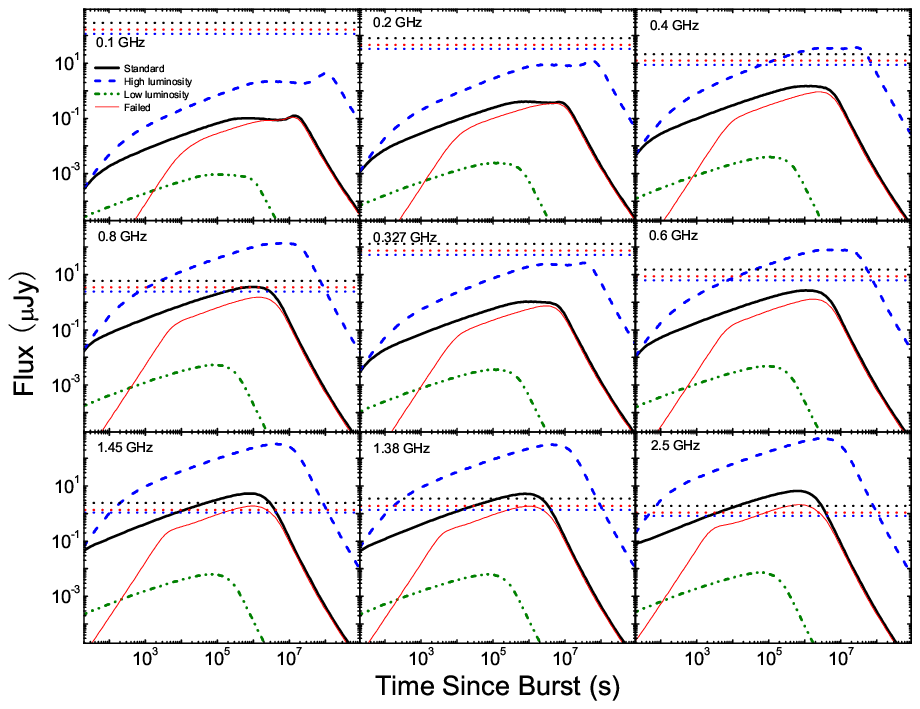}
   \caption{Radio flux density of GRBs at a redshift $z=5$ versus observation time $t$ in the observer frame at various frequencies within the FAST's window. Symbols are the same as in Fig. 1.}
   \label{Fig5}
   \end{figure}
   \begin{figure}
   \centering
\includegraphics[width=14.5cm, height=12cm,angle=0]{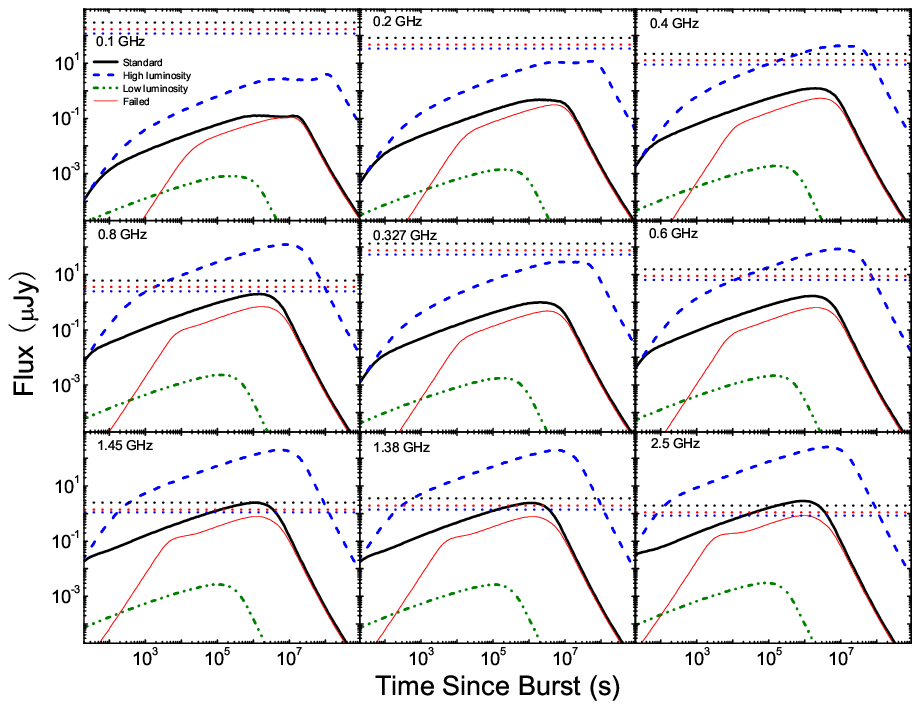}
   \caption{Radio flux density of GRBs at a redshift $z=10$ versus observation time $t$ in the observer frame at various frequencies within the FAST's window. Symbols are the same as in Fig. 1.}
   \label{Fig6}
   \end{figure}
   \begin{figure}
   \centering
 \includegraphics[width=14.5cm,height=12cm, angle=0]{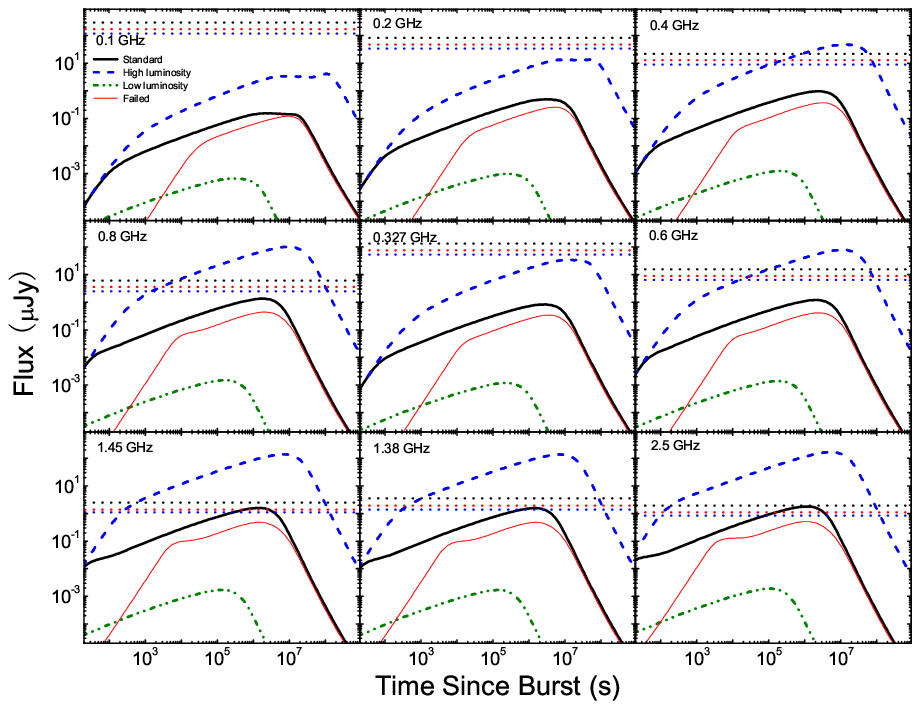}
   \caption{Radio flux density of GRBs at a redshift $z=15$ versus observation time $t$ in the observer frame at various frequencies within the FAST's window. Symbols are the same as in Fig. 1.}
   \label{Fig7}
   \end{figure}

 \subsection{Peak Spectra of Radio Afterglows in the FAST's Window}
To investigate the sensitivity of FAST's receiver at different frequencies, we plot the peak frequency against the peak flux density for the above-mentioned four types of bursts at different redshifts in Fig. 8. The data utilized here are extracted from the above calculations. Radio emission would be steeply cut off by the self-absorption effect at lower frequencies below several GHz. This frequency range covers the FAST's frequency bands of 70 MHZ-3 GHz, which causes the radio flux density $F$ to be a power-law function of $F\propto\nu^{2}$ (e.g. Sari, Piran \& Narayan 1998; Wu et al. 2005) if the observation frequency $\nu$ is less than the synchrotron self-absorption frequency $\nu_a$. Take the systematic temperature as $T_{sys}$=20 K, the bandwidth as $\Delta \nu=100$ MHz and the aperture efficiency as $\eta_A=0.65$ for FAST, we have plotted the 1$\sigma$ and 3$\sigma$ limiting flux density for a 10-minute integration time for comparison (see Fig. 8).
 \begin{figure}
   \centering
  \includegraphics[width=16cm,height=13cm, angle=0]{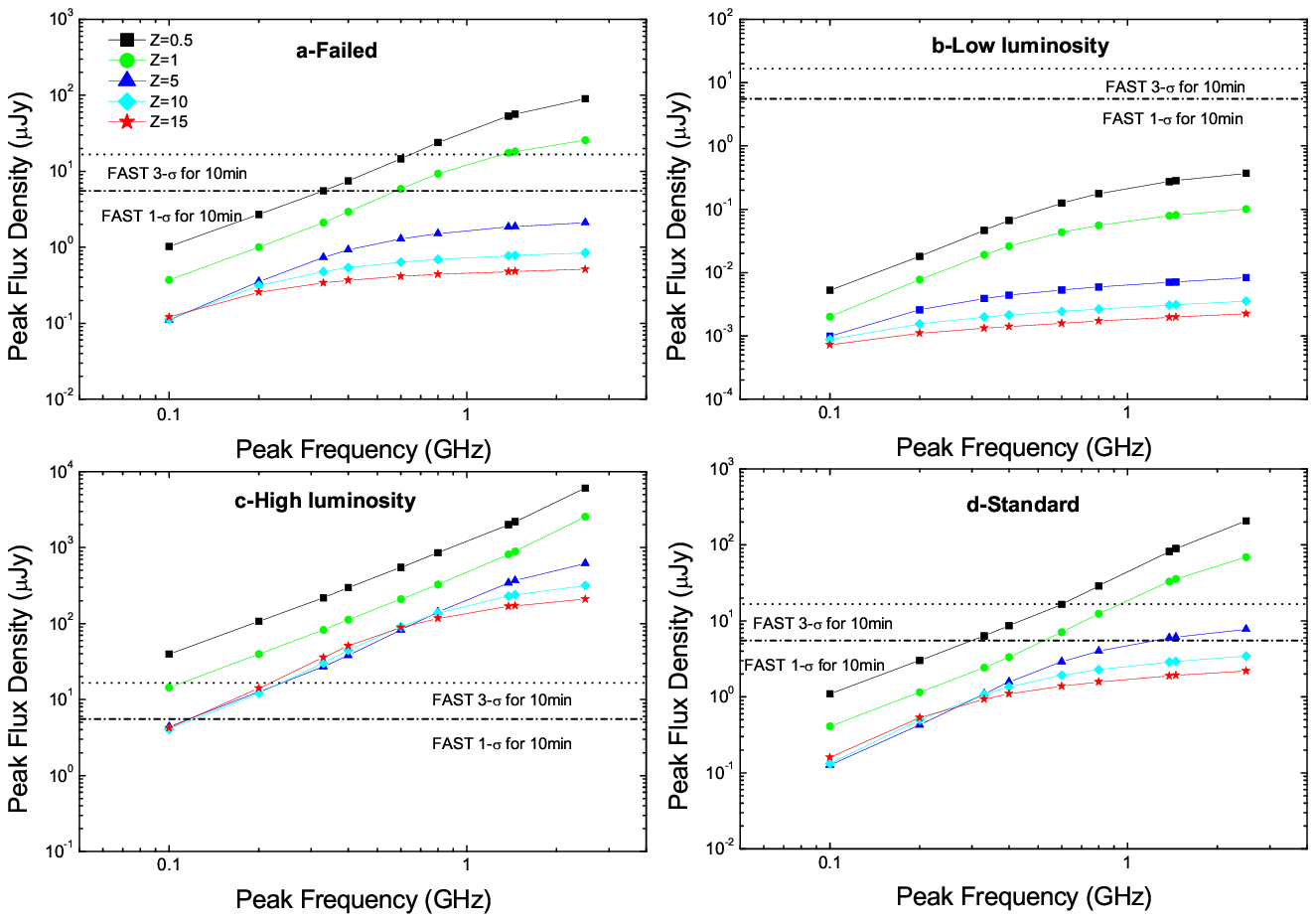}
   \caption{Peak radio flux density versus peak frequency for the failed, low luminosity, high luminosity and standard GRBs at five representative redshifts. The two horizontal lines from top to bottom respectively stand for 3$\sigma$ and 1$\sigma$ limiting flux density of FAST in a 10-minute integration time. The different symbols for various redshifts are given in the first panel. Note that a systematic temperature of $T_{sys}$=20K and a bandwidth of $\Delta\nu$=100 MHz as typical parameters for FAST have been used for our estimation here. See the text for details.}
   \label{Fig8}
   \end{figure}

We can see from Panel (a) in Fig. 8 that radio afterglows of failed bursts are detectable only for nearby sources with redshifts less than 5 and at relatively high frequencies. Panel (b) shows that the radio peaks of low luminosity bursts are far below the detection limits in all our cases and thus are difficult to detect by FAST. On the contrary, we see from Panel (c) that the peak flux densities of high luminosity GRBs are always above the detection limits, making them the best candidates for monitoring at almost any redshifts and frequencies except very high redshift at lower frequency of $\nu<$ 200 MHz. Panel (d) displays the radio peaks of the standard GRBs. They can be observed by FAST up to very high redshift of $z=$10 at higher frequency, very different from the failed bursts with relatively lower flux density at higher frequency, although they exhibit similar observational properties at lower redshifts in all the frequency bands of FAST. Another interesting phenomenon is that radio spectrum evolves with the cosmological redshift in the observer frame. In addition, the spectral shape of low luminosity bursts clearly differs from that of other classes of GRBs which hints they may be of distinct physical origin.

Recent numerical simulations by a few other authors have shown that sideways expansion, edge effect,\textbf{ even off-axis effect (Zhang et al. 2014a) of jets}, together with its microphysical process from ultra-relativistic to non-relativistic phase, may play an in-negligible role on the expected light curves and spectra of afterglows (van Eerten, Leventis, Meliani et al. 2010; van Eerten \& MacFadyen 2012). However, the results given by the above updated blastwave models do not significantly differ from those predicted by the generic afterglow model utilized in this paper, especially for the late-time radio afterglows, of which the peak flux density, the peak time and the post-peak decay are much comparable as a whole. Furthermore, it is worth pointing out that our numerical radio afterglows at 1.43 GHz is fairly consistent with those calculated with external shock model by Chandra \& Frail (2012).

\section{Discussion and Conclusion}
\label{sect:discussion}

It is interesting to notice that the fractions of high, low and medium isotropic energy GRBs in the pre-Swift era are 4\%, 16\% and 80\% respectively (Friedman \& Bloom 2005). As Swift/BAT is more sensitive to long-soft bursts than pre-Swift missions, the percentages are 32\%, 3\% and 65\% respectively for the high, low and medium isotropic energy GRBs in the Swift era. Obviously, the fraction of high luminosity GRBs detected by Swift/BAT is much larger than that of pre-Swift detectors, while the fraction of low luminosity GRBs is just the opposite. The on-going Swift satellite favors long bursts with higher redshifts on average. This naturally makes more and more super-long bursts observed with duration up to $10^4-10^6$s (Zhang et al. 2014b). How the radio afterglows of these GRBs will behave is a very important problem. We believe that their radio afterglows should be detectable to FAST. It will promote the study of early cosmology especially with the help of future FAST observations. This advantage is mainly attributed to the fact that long GRBs with high-redshifts are generally thought to be produced by higher luminosity sources, e.g. collapse of very massive stars.

Rhoads (1997) had pointed out that $\gamma$-ray radiation from some jetted GRBs can not be observed owing to relativistic beaming effects, but the corresponding late time afterglow emission is less beamed and can safely reach us. These lower frequency radiations are called as orphan afterglows, since they are not associated with any detectable GRBs. However, Huang, Dai \& Lu (2002) pointed out another possibility that orphan afterglows can also be produced by failed GRBs (or a dirty fireballs). They argued that the number of failed GRBs may be much larger than that of normal bursts. It is very difficult to distinguish the two different origins of orphan afterglows. The initial Lorentz factor of ejecta is a key parameter that makes the failed GRBs be different from other kinds of bursts in principle. Compactness limit was thought to be a robust method for estimating the initial Lorentz factor (Zou, Fan \& Piran 2011). Unfortunately, current Lorentz factor estimates are still controversial although extensive attempts had been made both theoretically and observationally (e.g. Zhang et al. 2007, 2011; Li 2010; Liang et al. 2010, 2011; Zou \& Piran 2010; Zou, Fan \& Piran 2011; Zhao, Li \& Bai 2011; Chang et al. 2012; Hascoet et al. 2013). It is helpful to discriminate between failed and standard GRBs from their radio afterglows because such low frequency emission can be observed for quite a long time (Huang, Dai \& Lu 2002). FAST may make important contribution in the aspect.

Note that contributions of host galaxies to radio fluxes have been neglected in our numerical calculations for simplification. This effect will add difficulties for afterglow observations with FAST and the detection rate of radio afterglows may be less optimistic than our current study (Li et al. 2014). On the other hand, the system temperature $T_{sys}$ is sensitively dependent on a variety of realistic factors and will be measured only after the radio telescope is built. As a single dish antenna, FAST will operate in a lower frequency range, i.e. from 70 MHz to 3 GHz in its first phase, and may extend to 8GHz in its second phase. The Low Frequency Array (LOFAR) covers relatively lower observation frequencies of 10-240 MHz (van Haarlem et al. 2013), partly overlapped with that of FAST. Thompson et al. (2007) gave the theoretical equation, $T_{sky}=60\lambda^{2.55}$ K, as the estimation of $T_{sys}$. It is illustrated that the LOFAR will be sky-noise dominated under 65 MHz, close to the FAST's lower frequency limit of 70 MHz. In contrast, at a frequency of 200 MHz, FAST's detection sensitivity is about one order of magnitude higher than LOFAR. In addition, we stress that the limiting sensitivity of FAST at $\nu >0.4$ GHz for 1-hour integration time is under 10 $\mu$Jy, which is much better than EVLA if FAST's instrumental noise is ideally controlled to be below the values adopted in this work. In this case, we deduce that FAST will be the most powerful new-generation radio telescope for studying radio afterglows. If the international Very-Long-Baseline Interferometry operation is applied, FAST would have more powerful capability of detecting radio emission from low luminosity GRBs or very distant (Chandra et al. 2010) GRBs, which should largely increase the detection rate of radio afterglows in the near future. Considering that GRBs with $E_{iso} = 10^{51} - 10^{53}$ erg at $z\geq20$ will be observed by the next generation instruments in near infrared and radio bands (Mesler et al. 2014), our results would be very valuable for making further survey plans of GRB radio afterglows with the upcoming radio telescopes such as FAST, SKA, and so on.

In conclusion, we summarize our major conclusions in the following.

\begin{enumerate}
\item[(1)] We presented a quantitative prediction for the detectability of GRB radio afterglows with FAST, based on the generic dynamical afterglow model. Our calculations are carried out for four kinds of bursts, i.e. failed GRBs, low luminosity GRBs, high luminosity GRBs and standard GRBs. We found that the radio afterglow detection rate sensitively depends on the model parameters of Lorentz factor and isotropic energy.
\item[(2)] We predicted that radio afterglows of all the above types of bursts except the low luminosity ones should be detected by FAST in wide ranges of time, frequency and redshift. The detectabilities descend in order for high luminosity, standard, failed and low luminosity GRBs.
\item[(3)] FAST is able to detect radio afterglows of GRBs at redshift up to $z\sim$10 or even more, which will be very helpful for the studies of GRB event rate and GRB cosmology.
\item[(4)] Radio afterglows of low luminosity GRBs will be detectable in the 2nd phase of FAST or if the integration time is extended enough.
\end{enumerate}

\normalem
\begin{acknowledgements}
We thank the anonymous referee for helpful comments and suggestions. We appreciate D. A. Frail and P. Chandra for delivering us their invaluable observation
data of radio afterglows. We acknowledge R. D. Nan, B. Zhang, X. F. Wu, Y. Z. Fan, C. S. Choi and H. Y. Chang for helpful discussions.
This work was supported by the National Basic Research Program of China (973 Programs, Grant No. 2014CB845800, 2012CB821800), the National Natural Science Foundation of China (Grant No. 11033002; 11263002; 11311140248) and Guizhou Natural Science Foundations (20134021; 20134005). SWK acknowledges support by China Postdoctoral science foundation under grant 2012M520382.

\end{acknowledgements}



\label{lastpage}
\end{document}